# High-Fidelity Fluid-Structure Interaction Simulations of Perforated Elastic Vortex Generators

**Karan Kakroo[a], Hamid Sadat[a,1]**

[a] Department of Mechanical Engineering, University of North Texas, Denton, Texas, USA

## Abstract

This study conducts a high-fidelity 2-way coupled Fluid Structure Interaction simulations, focusing on a novel perforated elastic vortex generator that is wall-mounted in an open channel with an incoming flow. The response of a perforated elastic vortex generator is investigated across a wide range of dimensionless parameters including dimensionless rigidity, mass ratios, Reynolds numbers, and porosity levels. Additional simulations for non-perforated elastic vortex generator are conducted for comparison and validation against available data. The findings demonstrate that a perforated elastic vortex generator can exhibit static configurations, lodging configurations, and vortex-induced vibration modes, depending on the dimensionless parameters. These configurations are similar to those observed for non-perforated elastic vortex generators, though the response values differ due to changes in the mechanical properties of the elastic vortex generator and the fluid loads acting on it as a result of perforation. Analysis of the perforated elastic vortex generator's natural frequencies shows that vortex-induced vibrations are triggered by the lock-in phenomenon associated with the second natural frequency of the vortex generator. Additionally, local flow dynamics are also studied by investigating vortical structures and velocity fields.

Keywords: Computational Fluid Dynamics; Fluid-Structure Interaction, Elastic Vortex Generator, Vortex Induced Vibration, Perforated Filament

## 1. Introduction

Vortex generators (VGs) play a pivotal role in intensifying mixing processes and enhancing heat transfer efficiency (HTE) [1–15] across a diverse array of industrial sectors, including but not limited to aerospace technology, automotive engineering, power generation plants and HVAC systems, chemical engineering, geothermal energy extraction, internally cooled gas turbine components, air-cooled solar energy harvesting systems, nuclear reactor designs, food processing facilities, and various other engineering applications [16–26]. For instance, in aerospace sector; VGs are utilized in thermal management systems to optimize heat exchangers used for cabin air conditioning and avionics cooling by enhancing the mixing and heat transfer rates. In automotive sector, they are installed in engine cooling systems such as radiator which leads to more effective heat dissipation. Additionally, they are used in air intake systems to improve the mixing of air and fuel before combustion which aids in better fuel atomization. In power generation plants and HVAC systems, VGs are used in arrays within air-cooled condensers, thus optimizing the cooling performance. In industries like oil and gas or chemical processing, VGs are implemented in separators and distillation columns to facilitate better separation of phases to increase overall efficiency. According to a recent market analysis of vortex generators, the projected growth of VGs indicates an annual expansion rate of 6.3%, with an anticipated market size nearing USD 34 million by 2026 [27]. In the realm of energy production, insights gleaned from the Environmental Impact Assessment (EIA) Annual Energy Review suggest that by integrating vortex generators to enhance the efficiency of existing heat exchangers in power plants, the United States stands to generate an additional 480 billion kWh of energy annually. This surplus energy output has the potential to satisfy forecasted demand increments until 2030, as projected by the EIA [28–32]. Thus, the pivotal role of vortex generators emerges as indispensable and potentially transformative across diverse sectors.

*Corresponding author
Email address: hamid.sadat@unt.edu

Extensive research has explored Rigid Vortex Generators (RVGs) to generate swirling motion in the flow to enhance mixing and heat transfer [26,33–40]. These studies have investigated different parameters such as different RVG shapes [41–49], number of RVG rows [50–52], arrangement of RVGs within the duct or in relation to each other [41–43,52], and attack angle of RVGs [44–47]. Regardless of shape of VG used, it has been concluded in these studies that winglets are generally more effective than wings [48,49], and delta winglets are superior to rectangular winglet pairs in terms of mixing intensification and heat transfer enhancement in laminar and transitional flow region, while curved trapezoidal delta winglets show better performance compared to conventional winglet VG in turbulent flow region [49,53,54]. Both inline and staggered arrangement are studied in detail in the literature, and it has been concluded that the staggered arrangement of winglets VGs yields higher mixing and heat transfer enhancement performance in a whole range of Reynolds number [52,55–57]. Additionally, it has been widely established in these studies that the longitudinal VGs are more efficient than transverse VGs as they disrupt the boundary layer and enhance the velocity gradient which yields in intensified mixing and better heat transfer rate/performance by exhibiting three mechanisms such as, secondary flow formation, developing boundary layer and intensifying turbulent intensity of fluid flow [6,16,48,49,53]. Moreover, studies exploring the influence of angle attack for various shapes of VGs on heat transfer and mixing intensification have concluded that an optimum attack angle lies in a range of $30^0$-$45^0$ for smallest pressure drop and highest mixing and thermal performance. Despite of all the different parameters investigated for the substantial enhancement of mixing and heat transfer, all these studies have revealed that the installation of RVGs in the system often comes with a penalty of increase in pressure drop [2,6,48,49,53,58–60]. Thus, substantial efforts have been made to reduce the pressure drop without compromising the mixing intensification and heat transfer enhancement.

Several studies have employed flexible or elastic vortex generators (EVGs) to reduce the pressure drop [13,33,59–70]. These studies have investigated the effect of different reconfiguration/bending rigidity of the EVG [71–75], different EVG shapes such as rectangular flat plates [33,59–61,65], circular or rectangular cylinders [46,76–78], flexible winglets [79–81], and flexible flags [63,69,82,83]. Additionally, the effect of number of EVGs (single and multiple/tandem) with different orientations such as clamped either at leading edge [84–86] or trailing edge [70,87,88], different EVG aspect ratios [59,89,90], and effect of different channel shapes such as circular pipes and rectangular channels [59,67,90–92] have been investigated in the literature as well. These studies have demonstrated that using an EVG leads to an intensified mixing and heat transfer with lower pressure drop compared to the RVGs [13,33,60,65,66,93]. However, these studies have concluded that more novel EVGs are still required to reduce the pressure drop even further while enhancing the mixing and heat transfer.

In this study, we explored using perforated EVGs (PEVGs) to achieve a more efficient VG with less pressure drop. High-fidelity 2-way coupled Fluid Structure Interaction (FSI) simulations are conducted to investigate the EVGs response for a wide range of dimensionless parameters obtained from governing equations for solid as well as fluid, including bending rigidity, mass ratio, Reynolds number, as well as porosity. Additional simulations are also conducted for non-perforated EVGs for comparison and validation against available experimental data. The focus was laid on the EVGs dynamics by investigating the response characteristics such as inclination angle; phase portrait; and response harmonics. The local flow is also studied by investigating the vortical structures and velocity field around the non-perforated and perforated EVGs. The considered VG shape in this study is a wall mounted filament which has been widely used in previous studies[53,61,64,71,72,75,89,94].

## 2. Geometry Information

Figure 1 shows the considered EVG in this study. The EVG is made up of an elastic material, having a length of L=0.1 m, thickness δ=0.008 m and the depth w=0.06 m. The porosity of the perforated EVG consisted of N=15 square holes of side a, which were distributed uniformly along the length (L) and depth of the structure (w), which resulted in an array of N= 15 (5 × 3) holes, as seen in Figure 1b. The selected range of porosities for this study were φ= [0, 0.1, 0.25, 0.35 and 0.5], which were controlled by adjusting the size of each pore. It is worth noting that the porosity of φ=0 represents non-perforated EVG. For the selected porosities (φ), a=0, 0.006325, 0.01, 0.0118321, 0.0141421 m, and the space between the holes was b=0.02, 0.013675, 0.01, 0.008167840, 0.0058579 m respectively, where "a" is the side of the square pore and "b" is the distance between pores.

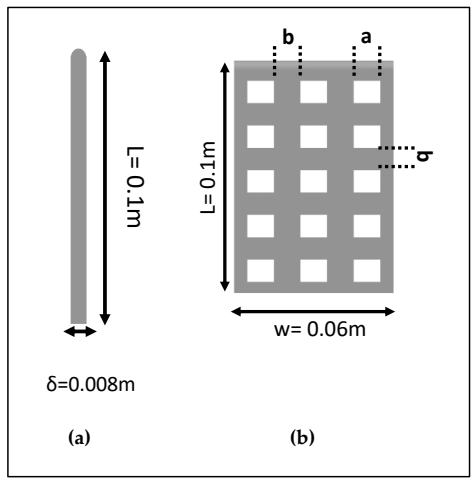

**Figure 1: Geometry of EVG.**

3. Computational Domain, Grid Information, and Boundary Conditions

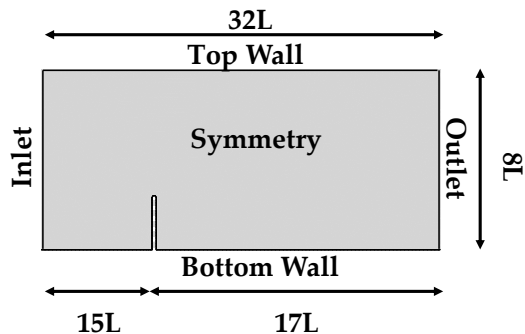

Figure 2: Computation domain.

The computational domain for this study is a rectangular box of [0,32L] × [0,8L] as shown in Figure 2. The distance from the inlet to the EVG is set to be 15L. The distance from the EVG to the outlet is set to be 17L. The EVG is clamped at the bottom wall as shown in Figure 2.

An unstructured grid for both the solid domain as well as fluid domain is generated for modeling fluid and structure interaction, shown by Figure 3a and Figure 3b respectively.

For the solid domain, an unstructured mesh of size 0.060166L was generated. A finer mesh having face sizing of 0.02L and refinement inside the pores was used to generate a high-quality mesh to capture the physics closely near the pores. This yielded approximately 53-64 cells along the depth, 92-106 cells along the length, and 3 cells along the thickness of the EVG, depending upon the porosity. Therefore, this resulted in a total number of nodes around 17K-27K and total number of elements around 51K-127K (shown by Figure 3a).

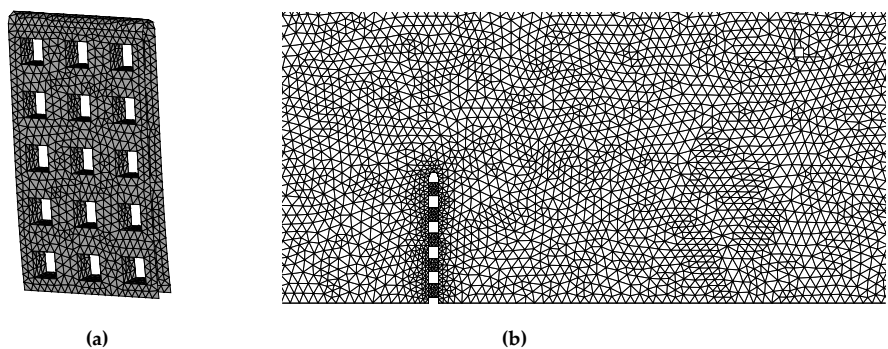

(a)                                          (b)

Figure 3: Grid Information.

An unstructured mesh of size 0.072288L was generated for the fluid domain. A face sizing of 0.03L (half the grid size of the domain) was used to generate a finer mesh on the FSI coupled surfaces as well as near the pores to capture the physics of the vortical structures in the near wake region. This yielded a total of 445 cells along the flow direction (209 cells from inlet to the EVG and 236 cells from EVG to outlet), 111 cells along the vertical direction and 8 cells along the depth far from the filament and approximately 23-37 cells near the filament, depending on porosity. Hence, this resulted in a total of about 2M unstructured cells for the fluid component. However, the grid number could potentially change slightly during the simulations since remeshing technique was used to update the fluid mesh during the simulation due to the dynamics of the EVG.

Moreover, to ensure if the results are insensitive/independent to the grid size, a grid verification study has been done by generating a set of grids with the refinement factor of $\sqrt{2}$ for both the solid as well as the fluid component. This resulted in total of five grid sets (G1 to G5) with the number of cells increasing from 2.9K to 37K cells and 58K to 2M cells for the solid and fluid domains respectively, from the coarsest to the finest grids, as discussed in later Section 6.1.

The boundary conditions used in this study are defined in Table 1.

**Table 1: Boundary conditions**

| Boundary Condition | Type | |
|---|---|---|
| | U | P |
| **Fluid component** | | |
| Inlet | $u = \vec{U}_{inlet}$ | $\frac{dP}{dn} = 0$ |
| Outlet | $\frac{d\vec{U}}{dn} = 0$ | P=P$_{constant}$ |
| Bottom wall | $\vec{U}=0$ | $\frac{dP}{dn} = 0$ |
| Top wall | $\vec{U}.\,n=0$ | $\frac{dP}{dn} = 0$ |
| Front and back | $\frac{d\vec{U}}{dn} = 0$ | $\frac{dP}{dn} = 0$ |
| **Solid component** | | |
| FSI Coupled faces | $\vec{U}_{fluid} = \vec{U}_{solid}$ | P$_{fluid}$ = P$_{solid}$ |
| Bottom face (clamped) | $\vec{U}=0$ | $\frac{dP}{dn} = 0$ |

For the fluid component, the inlet velocity boundary condition is used with values assigned corresponding to Reynolds number considered in this study. On the outlet, the outflow condition is implemented. A slip-wall condition is implemented on the top wall while a non-slip condition is implemented on the bottom wall. For the front and back, a symmetry condition has been imposed. Initially, the EVG is upright and has zero velocity. For the solid component, the bottom face is clamped and the EVG is vertically orientated.

### 4. Fluid and Structure solvers

Simulations are performed by solving fluid and solid governing equations using a partitioned two-way Fluid-Structure Interaction (FSI) coupling scheme [95]. The governing equations are

$$\frac{\partial \rho}{\partial t} + \nabla \cdot [\rho(v - v^m)] = 0 \qquad (1)$$

$$\rho \frac{\partial v}{\partial t} + \rho [(v - v^m) \cdot \nabla] v = \nabla \cdot \sigma + \rho F \qquad (2)$$

where, $\rho$ is the mass density, v is the (fluid or solid) particle velocity and $v^m$ is the grid point velocity (used to deform mesh to accommodate the deformation of the EVG) computed using a spring analogy where all point-to-point mesh connections are replaced with springs [96,97], $\sigma$ is the Cauchy stress tensor and F is the body force. The stress tensor for the fluid component is defined as follows:

$$\sigma = -pI + 2\mu S, \qquad (3)$$

where p is the thermodynamic pressure, $\mu$ is the absolute viscosity, I is the second rank unity tensor and S is the strain-rate tensor.

The Cauchy stress tensor for a linear elastic solid is defined as follows:

$$\sigma = 2\mu\varepsilon + \lambda\nabla \cdot UI, \qquad (4)$$

where $\mu$=E/2(1+ν) and $\lambda$= νE/(1+ ν)(1-2ν) are Lame's constants, and $\varepsilon$ is the Green-Lagrange strain tensor.

To discretize these equations, the least squares cell-based method was used for the gradient terms for special discretization and second order upwind discretization scheme was used for convective terms. A first order implicit transient formulation was used for all the cases done in this study. Moreover, a fully implicit coupled scheme was used for pressure-velocity coupling. .

### 5. Test Conditions and Validation Variables

To identify the dimensionless terms defining the physics of the EVG interaction with the flow and defining the test conditions, the governing equation for the motion of the EVG represented in Section 4 is simplified based on the Euler-Bernoulli beam theory assuming negligible shear stress resultant and written in the cantilever coordinate system as:

$$EI\left(\frac{\partial^4 X}{\partial s^4}\right) + M\frac{\partial^2 X}{\partial t^2} = \frac{F_p}{L^2} \qquad (5)$$

where, E is the flexural modulus of the structure, I is the second moment of area, $M = \frac{\rho_s A}{L}$ (A is the area of the EVG, A=Lδ), X denotes how much the structure bends or deflects from its original position at any point "s" along its length, and $\frac{F_p}{L^2}$ denotes the pressure loading per unit area. The Equation (5) can be rewritten in a dimensionless form as:

$$(\gamma)\frac{\partial^4 X^*}{\partial s^{*4}} + (\beta)\frac{\partial^2 X^*}{\partial t^{*2}} = \frac{F_p}{\rho_f U^2 L^2} \qquad (6)$$

where, $\gamma = \frac{EI}{\rho_f U^2 \infty L^4}$ 3 and $\beta = \frac{\rho_s \delta}{\rho_f L}$ are two dimensionless terms namely bending rigidity and mass ratio.

The term $\frac{F_D}{\rho_f U^2 L^2}$ represents the dimensionless force acting on EVG from the fluid which will be function of Reynolds number (Re), defined by $Re = \frac{\rho UL}{\mu}$, where $\mu$ is the dynamic viscosity of the fluid. Therefore, the structural behavior depends on at least three dimensionless terms including $\gamma$, $\beta$, and Re. In this study, simulations are conducted for $0.001 \le \gamma \le 5$, $0.05 \le \beta \le 6$, and $50 \le Re \le 800$. The simulations are also conducted for range of porosity $\phi = [0, 0.1, 0.25, 0.35 \text{ and } 0.5]$, defined as:

$$\phi \text{ (Porosity)} = \frac{Na^2}{Lw} \tag{7}$$

where, N is number of pores. This resulted in a total of 169 cases in this study.

To characterize the dynamic behavior of the EVG and validate results against available data in the literature, the inclination angle ($\theta$) of EVGs for all test conditions is investigated. The inclination angle is defined as the angle between the straight line connecting the tip and root of EVG and the incoming flow direction as shown in Figure 4.

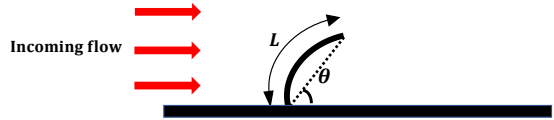

**Figure 4: A schematic diagram of EVG.**

The average inclination angle ($\theta_{ave}$) and the amplitude of the angular oscillation ($\theta_H$) are also investigated, which are defined as follows:

$$\theta_{ave} = \frac{1}{t_1 - t_0} \int_{t_0}^{t_1} \theta(t) dt \tag{8}$$

$$\theta_H = \{\max[\theta(t)] - \min[\theta(t)]\}_{t_0 \le t \le t_1} \tag{9}$$

where, the difference $t_1 - t_0$ is the time interval of oscillation of the EVG.

## 6. Results and discussions

### 6.1. Grid and Time-Step Verification
#### 6.1.1. Grid Verification

To ensure that the results are insensitive/independent of the grid size, a grid verification study is done by increasing the grid size several (four times) on both the solid as well as fluid component side by a factor of $\sqrt{2}$, resulting in total of five grid (fluid) sets ranging from G1= 2.019M cells which is the finest grid to G5=58K cells, which represents the coarsest grid.

The grid verification simulations are conducted for $\gamma = 0.1$, at a fixed $\beta = 1$ and Re = 400 condition, and at $\phi = 0.25$. At this condition, it is worth noting that EVG exhibited a static reconfiguration and there is no dynamic oscillation of the EVG observed (called VIV mode, as discussed later in Section 6.2). The predicted streamwise velocity profiles using different grid sizes at a distance L from the filament are investigated to evaluate the sensitivity of the results to the grid resolution, as can be seen in Figure 5.

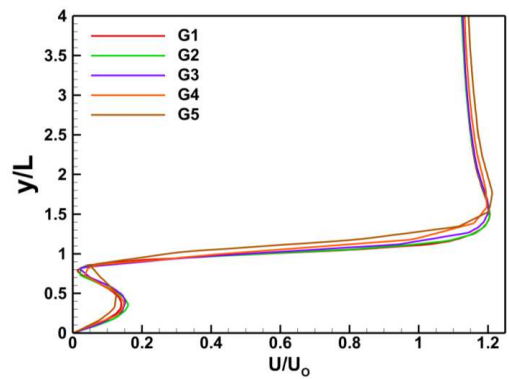

**Figure 5: Streamwise velocity profiles using different grid sizes.**

The results show all grids predict similar trends for the velocity profile in the wake of the filament. There is a significant flow velocity reduction in the wake of the EVG (y/L<1), while above that the flow velocity increases and almost becomes uniform. The results indicate a very close agreement between the predicted velocity field using grids G1 to G3, while the coarser grids G4 and G5 slightly show deviation from the rest. Hence, it is apparent from Figure 5 that the flow-field or the velocity profile does not change much with an increase in grid size from G3 to G1 and G3 should be sufficient to capture the flow field.

Even though the velocity profile/flow-field is predicted well with G3, it is imperative to check how fluid-structure interaction and consequently the predicted deformation for the EVG gets affected by change in grid size from G1 to G5. To check the grid resolution effect on the fluid-structure interaction, the sensitivity of the time history of the $\theta_{ave}$ to the grid size is investigated, as shown in Figure 6.

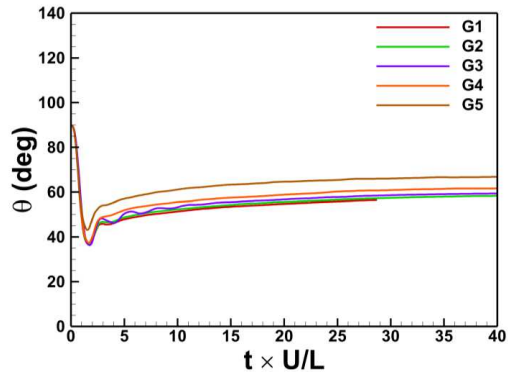

**Figure 6: Grid resolution effect on the fluid-structure interaction.**

It is apparent from Figure 6 that unlike the flow field which is fairly insensitive to the grids finer than G3, the EVG response eventually becomes insensitive to the grid size once the grid size reaches the G1 grid size, as there is slight deviation in $\theta_{ave}$=56.47⁰ of the EVG at the G1 grid size compared to $\theta_{ave}$=57.25⁰ at the G2 grid size. Hence for all the simulations conducted in this study, G1 grid size was chosen to ensure both flow and EVG response become insensitive with respect to the grid size. This grid is decomposed in 15 cores using parallel processing on a dedicated workstation, consuming approximately 183 K CPU hours for 169 cases investigated in this study.

### 6.1.2. Time-Step Verification

The dimensionless time step used in this study ranged from 0.04-0.2. which was non-dimensionalized by multiplying it by the ratio of flow inlet velocity to the length of the EVG. To ensure the results are independent of the time-step size, a time-step convergence study was conducted by varying the time-step several times. The time-step verification is conducted for similar test conditions using G1 grid size, used for grid verification study defined in Section 6.1.1.

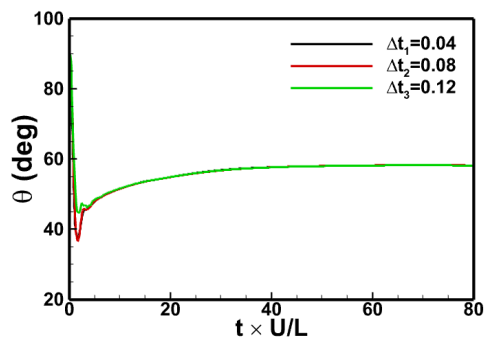

**Figure 7: Time-step Convergence study using different time steps.**

As shown by Figure 7, the results are insensitive to the range of dimensionless time-step used in this study, with only slight deviations in results for $\theta_{ave}$, such that $\theta_{ave} = 58.20^0$, $58.21^0$ and $58.17^0$ for $\Delta t_1 = 0.04$, $\Delta t_2 = 0.08$ and $\Delta t_3 = 0.12$ respectively. Therefore, the middle dimensionless time step $\Delta t_2 = 0.08$ was selected for majority of simulations done in this study to accurately capture the dynamic behavior of the EVG. It is important to note that for cases exhibiting dynamic oscillatory behavior of the EVG, the dimensionless time step was accordingly reduced to as low as $\Delta t_1 = 0.04$, to ensure there are at least 100 time-steps per cycle, ensuring adequate resolution of the physics and dynamic behavior of EVG. Additionally, for cases involving less deformation/bending of the EVG, the dimensionless time step was eventually increased to $\Delta t = 0.2$, to reach to converge condition faster.

### 6.2. Non-perforated EVG simulations

To validate the results, the simulations are first conducted for $\phi = 0$ (non-perforated condition), where data is available in the literature (Zhang et.al [64]). To compare the results against the perforated simulations, the effects of all three dimensionless parameters—dimensionless rigidity, mass ratio, and Reynolds number—are studied by varying one parameter while keeping the other two fixed.

The effect of $\gamma$ on the dynamic behavior of non-perforated EVG is investigated by fixing $\beta = 1.0$ and Re = 400. Figure 8 shows the variation of $\theta_{ave}$ and the $\theta_H$ as a function of $\gamma$.

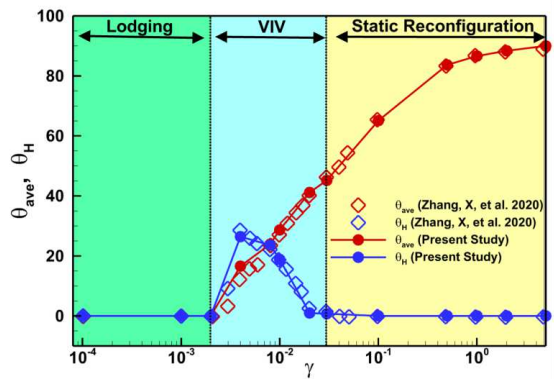

**Figure 8: Dynamic behavior of non-perforated EVG at different bending rigidities.**

From Figure 8, three different modes of dynamic behavior for non-perforated EVG can be observed, namely, lodging, VIV, and static reconfiguration. At low bending rigidity $\gamma \leq 0.002$ the lodging mode is observed (indicated by green color regime), where the non-perforated EVG bends and stays on the ground resulting in zero $\theta_{ave}$ and $\theta_H$. At large bending rigidity $\gamma \geq 0.03$, the static reconfiguration mode is observed (indicated by yellow color regime), where the non-perforated EVG bends slightly and stays in that position, thus has the $\theta_{ave}$ but zero $\theta_H$. At intermediate bending rigidity $0.002 < \gamma < 0.03$, the VIV (vortex-induced vibrations) mode is observed (indicated by blue color regime), where both the $\theta_H$ as well as the $\theta_{ave}$ are observed. This is because the vortices shed periodically from the tip of the non-perforated EVG, and the shedding frequency at Re = 400 is close to the second natural frequency of the EVG in that region of $\gamma$, resulting in the lock-in phenomenon, as described later in Section 6.6. In this regime, the $\theta_{ave}$ increases monotonically with increasing $\gamma$ since EVG becomes more rigid. The $\theta_H$ has a peak at $\gamma = 0.004$ and then decreases with increasing $\gamma$. Zhang et.al [64] also observed similar regimes and it is apparent that the results obtained in this study are in close agreement with their study, as shown in Figure 8.

The effect of $\beta$ on the dynamic behavior of the non-perforated EVG is investigated by fixing $\gamma = 0.01$ and Re = 400, as shown in Figure 9. It is worth noting that as per the definition of $\beta$, the larger the $\beta$, the larger the density of solid, meaning that the structure would be more heavy and thus difficult to retract back to its upright position.

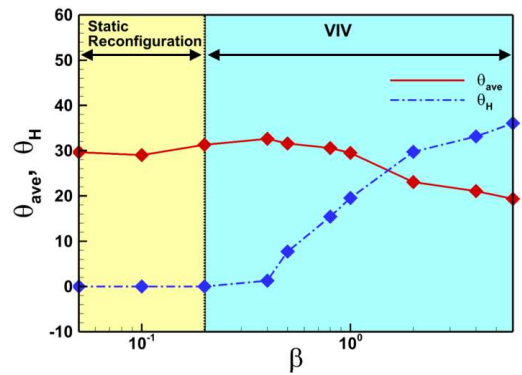

**Figure 9: Dynamic behavior of non-perforated EVG at different mass ratios.**

From Figure 9, the static reconfiguration mode is observed for $\beta \leq 0.2$, where the non-perforated EVG bends slightly and stays in that position. This is because the buoyancy force is larger compared to the gravity, hence counteracting the bending moment on EVGs induced by the fluid flows. The $\theta_{ave}$ in this region is approximately close to $30^0$ for the considered $\beta$ values. The $\theta_H$ is zero in this regime as the non-perforated EVG bends with zero oscillation. However, with increase in value of $\beta > 0.2$, the EVG starts exhibiting the VIV mode, as the vortex shedding frequency is close to the second natural frequency in that range of $\beta$, as discussed later in Section 6.6. Within this range, $\theta_H$ increases monotonically with increasing $\beta$ due to the lock-in phenomenon as discussed later, while opposite trend is observed for $\theta_{ave}$. $\theta_{ave}$ and $\theta_H$ are expected to eventually reach to nearly zero at very large $\beta$ values demonstrating the lodging configuration due to the heaviness of the EVG.

The effect of Re on the dynamic behavior of the non-perforated EVG is investigated by fixing $\gamma = 0.01$ and $\beta = 1.0$, as shown in Figure 10.

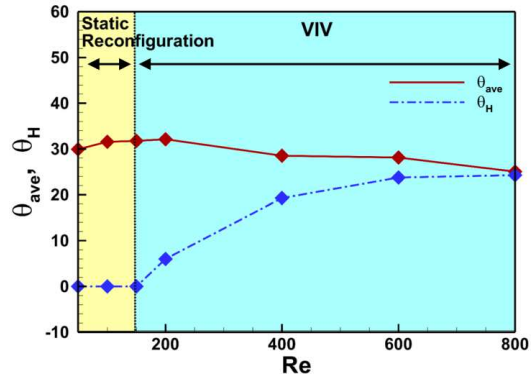

**Figure 10: Dynamic behavior of non-perforated EVG at different Reynolds numbers.**

Figure 10 shows that for Re ≤ 150, static reconfiguration is observed as the non-perforated EVG bends with zero $\theta_H$. With an increase in Re > 150, the flow becomes unsteady with a periodic vortex shedding at the tip of the EVG which leads to a transition from static reconfiguration to a VIV configuration. Within this range, $\theta_H$ increases with increasing Re, while $\theta_{ave}$ decreases slowly by increasing Re due to the larger drag force on the EVG at higher Re number.

### 6.3. Perforated EVG simulations

The perforated EVG simulations are conducted for various $\phi = [0.1, 0.25, 0.35, 0.5]$ and the effects of all three dimensionless parameters—dimensionless rigidity, mass ratio, and Reynolds number—are studied by varying one parameter while keeping the other two fixed. The results are compared with the results obtained in the previous Section 6.2 for non-perforated EVG.

The effect of $\gamma$ on the dynamic behavior of the perforated EVG is investigated by fixing $\beta$ = 1.0 and Re = 400, as shown in Figure 11. It should be noted the $\gamma$ for perforated EVG is defined based on the second moment of area (I) at the base of the EVG since I is no longer constant along the EVG due to the presence of pores.

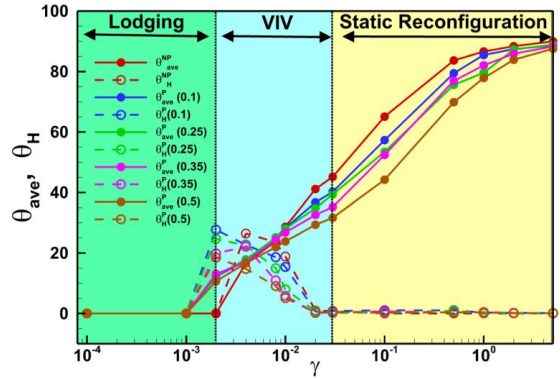

**Figure 11: Dynamic behavior of perforated EVG at different bending rigidities.**

It is apparent from Figure 11, for different porosities, three different modes namely lodging, VIV and static reconfiguration are still observed, identical to what was observed for non-perforated EVG, however the VIV mode is shifted towards the lower $\gamma$ values. The shift is apparent from the Figure 11, as the $\theta_H$ is non-zero between $0.001 < \gamma < 0.02$ instead of $0.002 < \gamma < 0.03$ which is the VIV region for the non-perforated cases, as colored by blue. The reason for this shift is the change of the natural frequency of the perforated structures as discussed later in Section 6.6, which also shifts the peak of $\theta_H$ to lower $\gamma$ values for perforated EVGs. Moreover, the values of $\theta_{ave}$ for perforated EVGs are often smaller than those for the non-perforated cases. This could be due to the effects of perforations on the loads (drag and lift) on EVGs and/or the change of the mechanical property of perforated EVGs as the moment of area along perforated EVGs varies. This will be discussed in detail in Section 6.3.1 and Section 6.3.2.

The effect of $\beta$ on the dynamic behavior of the perforated EVG is investigated by fixing $\gamma$ = 0.01 and Re = 400, as shown in Figure 12. From Figure 12, for all perforated EVG, it is observed that the static reconfiguration mode is shifted from $\beta \leq 0.2$ to $\beta \leq 0.5$ when compared to non-perforated EVG. With an increasing in values of $\beta > 0.5$, perforated EVGs start exhibiting VIV mode. Within this region, at $\phi$ =0.1, $\theta_H$ increases and $\theta_{ave}$ decreases with increasing $\beta$, identical to what is observed for non-perforated self-excited EVG. Interestingly for $\phi$ = [0.25, 0.35, 0.5], the trend observed for $\theta_{ave}$ is still similar to that for non-perforated cases, while the $\theta_H$ increases only up to $\beta$ = 2, and then lodging mode is observed, where the $\theta_H$ is zero. The lodging mode observed at higher $\phi$ values is because of not enough lift for the EVG at high level of porosity to counteract the large bending moment induced by the weight of structure at high $\beta$ values. Figure 12 also shows the $\theta_{ave}$ often reduces with an increase of the perforation which could be due to the effects of perforations on the loads on EVGs and/or the change of the mechanical property of perforated EVGs, which will be discussed in Section 6.3.1 and Section 6.3.2.

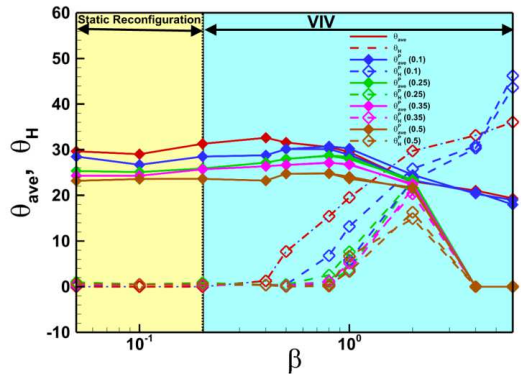

**Figure 12: Dynamic behavior of perforated EVG at different mass ratios.**

The effect of Re of maximum 800 on the dynamic behavior of the perforated EVG is investigated by fixing $\gamma = 0.01$ and $\beta = 1.0$, as shown in Figure 13. The results plotted in Figure 13 show similar trends for all porosities i.e. $\theta_H$ increases with increasing Re, while $\theta_{ave}$ decreases slowly by increasing Re due to the larger drag force on EVG at higher Re number. However, values of both $\theta_{ave}$ and $\theta_H$ reduce by increasing porosity.

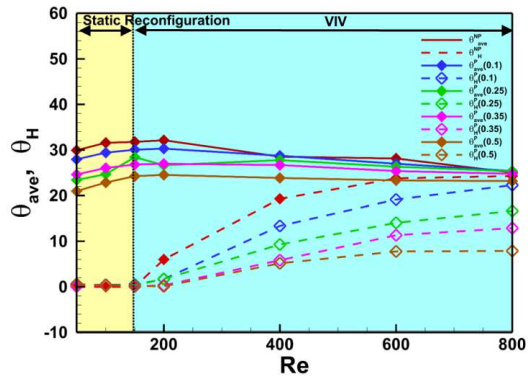

**Figure 13: Dynamic behavior of perforated EVG at different Reynolds numbers.**

Overall, Figures 11, 12, and 13 suggest that perforated EVGs can exhibit similar modes to those observed in non-perforated EVGs, though different regions for these modes may shift slightly and there are differences in inclination angle values for perforated and non-perforated EVGs. The sources of these differences are explored in the following sections.

### 6.3.1. Effect of perforation on mechanical property of EVGs

To analyze the trends observed in previous figures, and define whether the difference in perforated and non-perforated deformations reported in previous figures are due to the effect of pores on the mechanical property of EVGs, several solid mechanic simulations were conducted first to adjust the flexural modulus (E) of perforated EVGs to ensure perforated EVGs with non-uniform second moment of area and non-perforated EVGs with uniform second moment of area result in the same structural response under the same load. In those simulations, the force was applied at the tip of the EVG for two different magnitudes (0.12 N and 0.24 N) while the other end (base) of the EVG was clamped. The obtained flexural modulus for perforated cases giving the same deformation as non-perforated case for $\gamma = 0.1$ are shown in Table 2.

**Table 2: Effect of perforation on mechanical property of EVGs**

| Porosity | Modulus of the structure (E) | Force (N) | Average inclination angle ($\theta_{ave}$) |
|----------|------------------------------|-----------|--------------------------------------------|
| $\phi$=0 | 937500 | F=0.12 | 81.36827 |
|          |        | F=0.24 | 73.46998 |
| $\phi$=0.25 | 1516622.179 | F=0.12 | 82.26645 |
|             |             | F=0.24 | 74.82964 |
| $\phi$=0.5 | 2644726.486 | F=0.12 | 80.83791 |
|            |             | F=0.24 | 72.54316 |

From Table 2, it is apparent that the perforated cases with $\phi = 0.25$ and $\phi = 0.5$ require 1.6 and 2.8 times larger flexural modulus than the one for the non-perforated EVGs, respectively, to compensate the second moment of area reduction by pores and deform similarly as non-perforated cases.

In the literature [94], an equation for equivalent bending stiffness ($EI_P$) for perforated filament clamped at one end has been developed theoretically as follows:

$$EI_P = KEI_{NP} \tag{10}$$

where

$$K = \frac{(N+1)B(N^2+2N+B^2)}{(1-B^2+B^3)N^3+3BN^2+(3+2B-3B^2+B^3)B^2N+B^3} \tag{11}$$

Here, $B = \frac{b}{b+a}$; N is number of holes, $EI_{NP}$ is bending stiffness for non-perforated structures, $EI_P$ is bending stiffness for perforated structures. Equation 10 suggests that the bending stiffness for the perforated filaments reduces by a factor K due to the presence of pores. Therefore, the flexural modulus of perforated cases should be increased by the factor 1/K to have a similar deformation for perforated and non-perforated filaments. For $\phi$=0.25, the 1/K for N=15 and a=0.01 and b=0.01 used in this study is 1.618, and for $\phi$=0.5 with a=0.0141421 and b= 0.0058579 is 2.821. These values agree well with the ratio of the obtained E values for perforated and non-perforated EVGs reported in Table 2.

Using the obtained flexural modulus in Table 2, simulations for perforated EVGs with porosity of $\phi$ =[0.25, 0.5] were reconducted for the entire range of $\gamma$, fixing $\beta = 1.0$ and Re = 400. With the updated bending stiffness, the values for $\theta_{ave}$ do not decrease significantly by increasing perforation anymore, which was the trend observed in Figure 11. Instead, the $\theta_{ave}$ values are close to the one for non-perforated cases in the static mode and larger in the VIV region. The amplitude of oscillation also seems to constantly reduce by increasing the porosity unlike the one shown in Figure 11. Since non-perforated and perforated EVGs with similar mechanical properties exhibit different responses, it can be concluded that perforation affects the fluid loads on the EVG, as explored in the next Section 6.3.2.

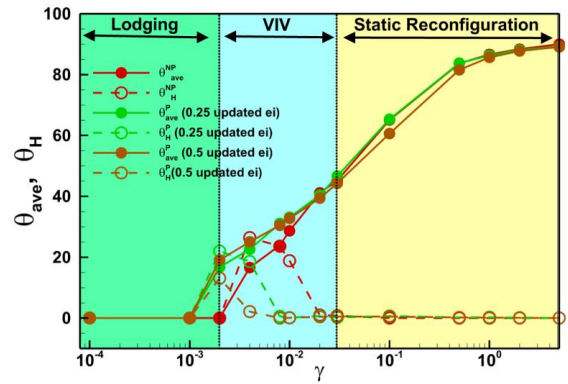

**Figure 14: Dynamic behavior of perforated EVG at different bending rigidities with adjusted bending stiffness.**

### 6.3.2. Effect of perforation on fluid loads

To evaluate the effect of perforation on fluid loads, drag and lift forces for the cases shown in Figure 14 are analyzed and reported in Table 3. For these cases, any differences in results are no longer attributed to variations in mechanical properties and should be solely because of the fluid loads.

**Table 3: Effect of perforation on fluid loads**

| Porosity | Bending rigidity ($\gamma$) | Drag (N) | Lift (N) |
|---|---|---|---|
| $\phi$=0 | 0.1 | 0.912 | -0.2736 |
| $\phi$ =0.25 updated EI | (Static mode) | 0.792 | -0.219 |
| $\phi$ =0.5 updated EI | | 0.585 | -0.165 |
| $\phi$=0 | 0.03 | 0.5244 | -0.3312 |
| $\phi$ =0.25 updated EI | (Static mode) | 0.444 | -0.2466 |
| $\phi$ =0.5 updated EI | | 0.3005 | -0.1397 |
| $\phi$=0 | 0.004 | 0.179897 | -0.098216 |
| $\phi$ =0.25 updated EI | (VIV mode) | 0.199264 | -0.07773 |
| φ =0.5 updated EI | | 0.165317 | -0.049331 |
| φ=0 | 0.002 | - | - |
| φ =0.25 updated EI | (VIV mode) | 0.1557936 | -0.0411881 |
| φ =0.5 updated EI | | 0.1335782 | -0.0230144 |

From Table 3, it is apparent that with increasing porosity, there is a reduction in drag values and downward force (negative lift) for EVGs experiencing static configuration, suggesting less pressure drop for perforated EVG. Although the loads are reduced, interestingly the deformation of perforated and non-perforated EVGs are relatively similar as shown in Figure 14. This is because perforation changes the distribution of fluid forces on EVGs. From the pressure contours plotted in Figure 15, it is apparent that the high-pressure region on the windward side as well as the low-pressure region on the leeward side are limited to the middle region of the perforated EVG while non-perforated EVG shows a more uniform distribution. The change of the pressure distribution shifts the action point of pressure forces toward the tip of the perforated EVG. This increases the bending moment, resulting in a similar inclination angle or bending for both perforated and non-perforated EVGs, despite the smaller loads on the perforated EVGs.

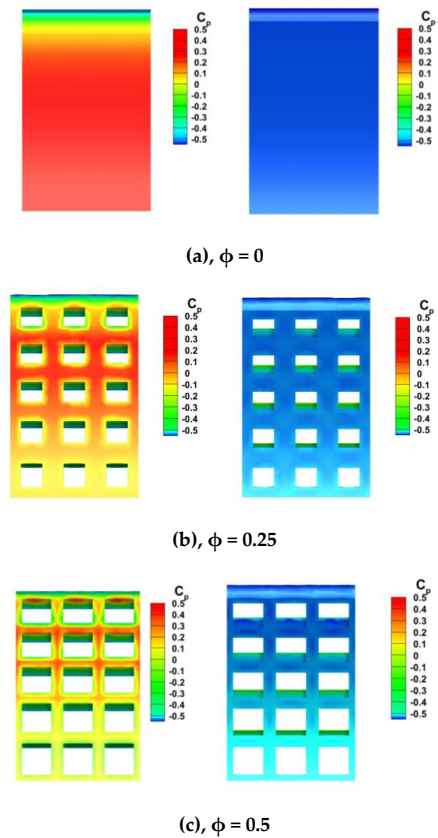

(a), φ = 0

(b), φ = 0.25

(c), φ = 0.5

Figure 15: Pressure contours at $\gamma = 0.1$ of both non-perforated and perforated EVG with adjusted bending stiffness.

Table 3 also shows a reduction in loads (or less pressure drop) when increasing porosity from 0.25 to 0.5 for cases experiencing VIV, while both cases exhibit a very similar mean inclination angle, as shown in Figure 14. This suggests that the distribution of loads is influenced by the level of perforation, leading to the same mean inclination angle, despite differences in loads. Additionally, Table 3 interestingly reveals that drag force increases when porosity is raised from 0 to 0.25, indicating that under certain conditions, porosity might increase drag force. This is possibly due to the significant alteration of the effective velocity caused by bending of the structure. This phenomenon of drag increase at certain porosity conditions has also been reported in the limited studies [94,98]. Further investigation is needed to fully understand these interactions, which is beyond the scope of the current study.

### 6.4. Dynamic behavior of self-excited EVG

To further investigate the dynamics of EVGs in VIV region, the time history; phase portrait; and the Fast Fourier Transform (FFT) of θ are studied. The effect of $\gamma$ on the dynamic behavior of the self-excited non-perforated and perforated EVG is investigated by fixing $\beta = 1.0$ and Re = 400. Two periodic oscillation cases in regular VIV mode are considered having $\gamma = 0.01$ and $\gamma = 0.004$. The case with $\gamma = 0.004$ was selected as it corresponds to the maximum $\theta_H$, as indicated by the peak in Figure 11 for the non-perforated EVG. The $\gamma = 0.01$ condition was chosen because it represents the furthest condition from $\gamma = 0.004$ where VIV is still observed with a large amplitude, providing two distinct cases for comparison.

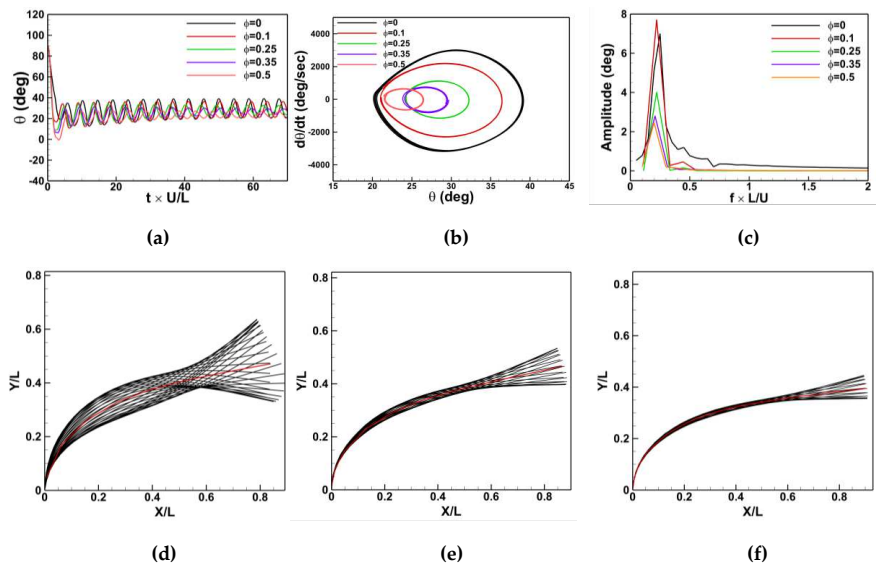

**Figure 16:** Dynamic behavior of a self-excited EVG at $\gamma = 0.01$ with fixed $\beta = 1.0$ and Re = 400, (a) Time history of the inclination angle, (b) Phase portrait, (c) Fast Fourier Transform (FFT), and d), (e) and (f) showing the time history of the bending pattern of self-excited EVG for $\phi = 0$, $\phi = 0.25$ and $\phi = 0.5$ respectively.

Figure 16a exhibits the bending of the self-excited non-perforated and perforated EVG over time at $\gamma = 0.01$. It is apparent that the non-perforated and perforated self-excited EVG bends initially from its upright position ($\theta = 90$ degrees) showing significant initial bending and then a dynamic oscillatory behavior with an amplitude of oscillation (which depicts the VIV mode) is observed with time passing by. For a non-perforated self-excited EVG at $\phi = 0$, the structure bends from its initial upright position ($\theta = 90$ degrees) to an angle of almost 20 degrees due to the fluid forces acting on the non-perforated EVG before showing a dynamic oscillatory behavior due to the coherent vortex shedding discussed in Section 6.5. With the increase in porosity, $\theta$ for self-excited perforated EVG becomes slightly lower compared to a non-perforated self-excited EVG, due to their lower second moment of area due to perforation.

Figure 16b exhibits the phase diagram plot for $\gamma = 0.01$. The elliptical shape phase diagram indicates a stable dynamic system. The size of the ellipse is largest and widest for non-perforated self-excited EVG compared to the size of ellipse observed for perforated self-excited EVG at all porosities. This corresponds to the larger amplitude of oscillation observed in Figure 16a for non-perforated EVG compared to that for all ranges of porosity. For non-perforated case, the elliptical shape is also slightly asymmetric and different for positive and negative rate of the inclination angle revealing that the fluid loads on non-perforated EVG changes during the acceleration and deacceleration phases in each cycle.

Figure 16c exhibits the Fast Fourier Transform plot showing a first harmonic response at the frequency corresponding to the main period of oscillation observed in Figure 16a. There also exists, a second peak (second harmonic) as well as third peak (third harmonic) for non-perforated self-excited EVG, however they are not as prominent as the first peak. With an increase in porosity, the first peak reduces, and the frequency appears to shift compared to the trend observed for non-perforated self-excited EVG. The shifting of frequency is due to the change in the oscillating frequency for perforated self-excited EVG. Identical to non-perforated self-excited EVG, no prominent second or third peak are observed for perforated cases.

Figure 16d, e and f show the stationary sinusoidal dynamic pattern with the average oscillatory profile indicated by red line for porosity of $\phi = 0$, $\phi = 0.25$ and $\phi = 0.5$ respectively. It is apparent from these figures that clamped EVG experiences the second mode of oscillation, as the shape is characterized by a sinusoidal wave along the length of the EVG for both non perforated and perforated EVGs [99].

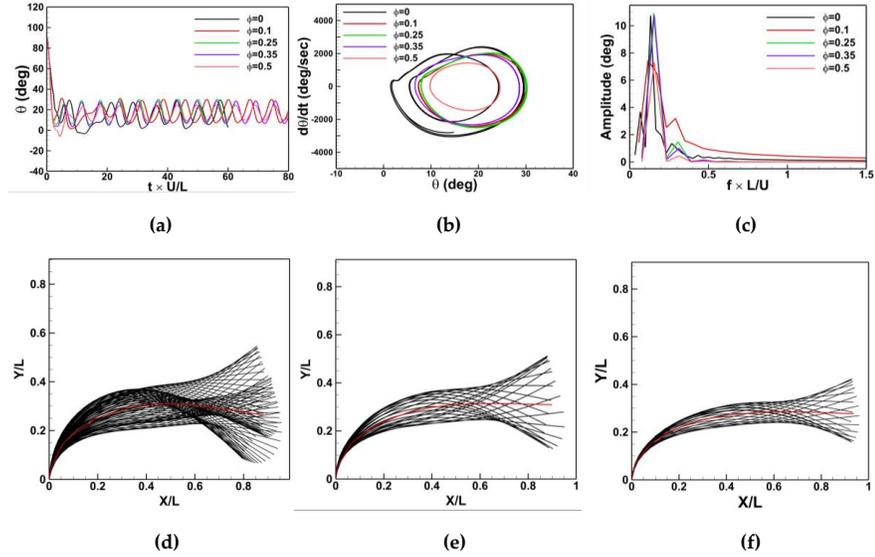

**Figure 17:** Dynamic behavior of a self-excited EVG at $\gamma = 0.004$ with fixed $\beta = 1.0$ and Re = 400, (a) Time history of the inclination angle, (b) Phase portrait, (c) Fast Fourier Transform (FFT), and (d), (e) and (f) showing the time history of the bending pattern of self-excited EVG for $\phi = 0$, $\phi = 0.25$ and $\phi = 0.5$ respectively.

Figure 17 shows the time history; phase portrait; and FFT of inclination angle of the self-excited non-perorated and perforated EVG at $\gamma = 0.004$. It is evident from Figure 17a, that the amplitude of the oscillation for non-perforated is more than the one observed for $\gamma = 0.01$. In addition, the trend observed for non-perforated self-excited EVG at $\gamma = 0.004$ is non-linear compared to the linear trend observed for non-perforated self-excited EVG at $\gamma = 0.01$. With the increase in porosity, the oscillatory trend becomes linear. The amplitude of the oscillation of perforated EVGs is often smaller than the amplitude of oscillation for non-perforated self-excited EVG, which is consistent with findings from Figure 11.

Figure 17b indicates two dynamic cycles for non-perforated EVG (indicated by a dip in the black color loop for $\phi = 0$), unlike for $\gamma = 0.01$. With an increase in porosity ($\phi = 0.1, 0.25, 0.35, 0.5$), EVGs only show one dynamic cycle. This is also apparent from Figure 17d, which shows the non-linear oscillatory bending pattern for non-perforated case which has two dynamic cycles, one in which the perforated EVG bends to a certain extent and the other dynamic cycle where it bends even further, thus exhibiting two different dynamic bending cycles unlike the one for perforated cases in Figures 17e and 17f.

From Figure 17c, it is noticeable that non-perforated self-excited perforated EVG shows a first harmonic at the frequency almost half of the of the frequency predicted at $\gamma = 0.01$, suggesting the increase of rigidity decreases the natural frequency of EVGs as discussed later. Moreover, there is a half harmonic observed before the first peak, which is in consistent with the trend observed in Figure 17a and 17b. Higher harmonics peaks are also observed but are not prominent as the first harmonic peak. With an increase in porosity, the first and higher harmonic amplitudes slightly decrease with increase in porosity and the half harmonics vanishes, which is again consistent with the trend observed in Figure 17a and 17b.

To understand the effect of $\beta$ on the dynamic behavior of the self-excited non-perforated and perforated EVG, Figure 18 shows the time history; phase portrait; and FFT of inclination angle at $\beta$ = 2. The value $\beta$ = 2 was selected based on the observation that for porosities $\phi$= [0.1, 0.25, 0.35, 0.5], the trends of $\theta_{ave}$ and $\theta_{H}$ for perforated cases closely resemble those of non-perforated cases up to this $\beta$ value. Beyond $\beta$ = 2, however, a transition to a lodging mode is observed for $\phi$=[0.25,0.35,0.5], where $\theta_{H}$ reaches zero. This transition presents an intriguing case for further investigation. Additionally, it is pertinent to compare these trends with those observed for $\beta$=1, as illustrated in Figure 16.

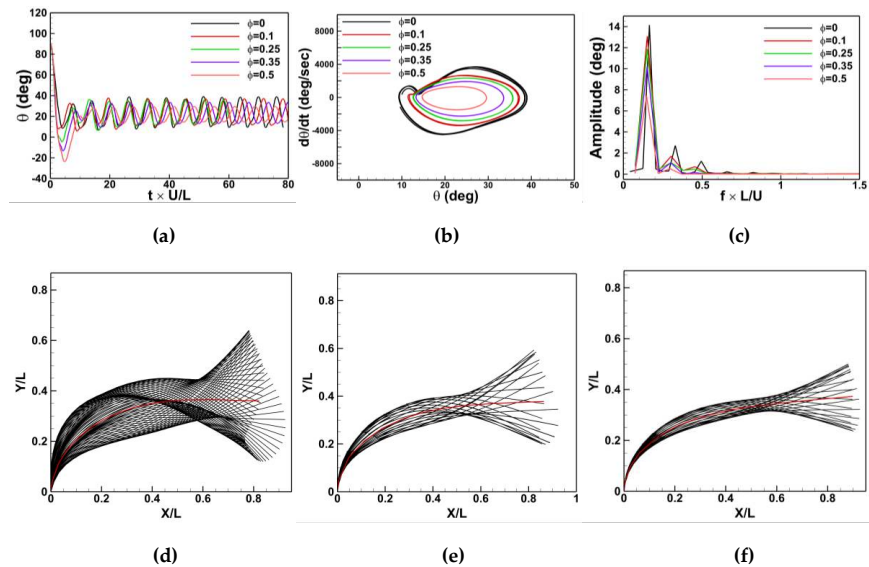

**Figure 18: Dynamic behavior of a self-excited EVG at $\beta$ = 2 with fixed $\gamma$ = 0.01 and Re = 400, (a) Time history of the inclination angle, (b) Phase portrait, (c) Fast Fourier Transform (FFT), and (d), (e) and (f) showing the time history of the bending pattern of self-excited EVG for $\phi$ = 0, $\phi$ = 0.25 and $\phi$ = 0.5 respectively.**

Figure 18a and 18b exhibit a linear oscillatory pattern in the dynamic behavior of non-perforated self-excited EVG at $\beta$ = 2, identical to the linear sinusoidal behavior observed at $\beta$ = 1 in Figure 16. The size of the loop for non-perforated self-excited EVG is largest as the amplitude of the oscillation is largest in this case compared to the perforated self-excited EVG at all porosities. With an increase in porosity, the dynamic cycle still exhibits a stationary linear sinusoidal pattern, which is identical to the linear trend observed for all porosities at $\beta$ = 1 for self-excited perforated EVG.

Figure 18c exhibits FFT showing that for non-perforated self-excited perforated EVG, there is a first harmonic occurring at lower frequency compared to the one for $\beta$ = 1, indicating that the increase of the mass ratio decreases the natural frequency of EVGs. There also exists, a second harmonic as well as third harmonic for non-perforated self-excited EVG, indicating that the dynamic response of the non-perforated EVG at $\beta = 2$ is highly non-linear unlike at $\beta = 1$. With an increase in porosity, the first peak reduces, and the higher harmonics becomes less pronounced, indicating more linear behavior. In addition, the frequency appears to shift compared to the trend observed for non-perforated self-excited EVG due to the change in natural frequency of the self-excited perforated EVG as discussed later in Section 6.6.

To examine the impact of Re on the dynamic behavior of both self-excited non-perforated and perforated EVGs, the time history, phase portrait, and FFT of the inclination angle are plotted for Re = 600 in Figure 19 and compared with the trends observed for Re = 400 in Figure 16. Figure 19a and b illustrate very similar dynamic behavior for the EVG to the one observed at Re = 400. From Figure 19c, it is observed that for non-perforated self-excited perforated EVG, there is a first harmonic occurring around a frequency slightly smaller than the one observed at Re = 400, both close to the Strouhal number of 0.2, reported in the literature as the Karman vortex shedding frequency at this range of Reynolds number. There also exists, a second harmonic as well as third harmonic for non-perforated self-excited EVG, however they are not as prominent as the first harmonic. With an increase in porosity, the first harmonic reduces and the higher harmonics diminishes, exhibiting more linear behavior while the frequency appears to shift compared to the trend observed for non-perforated self-excited EVG. Figure 19 (d), (e) and (f) shows the time history of the EVG profile with the average oscillatory profile indicated by red line for porosity of $\phi = 0$, $\phi = 0.25$ and $\phi = 0.5$ respectively. These figures clearly show that the EVG undergoes a second mode of oscillation similar to that observed at Re = 400, with the amplitude of oscillation decreasing as porosity increases.

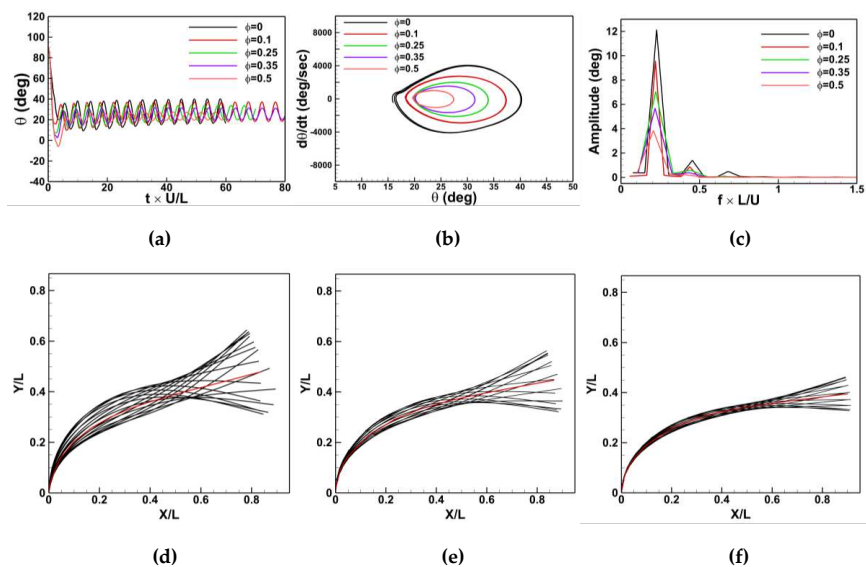

**Figure 19: Dynamic behavior of a self-excited EVG at $Re = 600$ with fixed $\gamma = 0.01$ and $\beta = 1$, (a) Time history of the inclination angle, (b) Phase portrait, (c) Fast Fourier Transform (FFT), and (d), (e) and (f) showing the time history of the bending pattern of self-excited EVG for $\phi = 0$, $\phi = 0.25$ and $\phi = 0.5$ respectively.**

### 6.5. Local flow study

The local flow information of self-excited non-perforated and perforated EVG is studied by investigating the vortical structures (shown by the dimensionless vorticity plot) around the self-excited EVGs. Figures 20 and 21 show the local flow plotted at several instances for self-excited EVG at $\gamma$ = 0.004, where the oscillation amplitude is maximum (as shown in Figure 11), twelve instances for $\phi$ = 0 are plotted, highlighting the non-linear trend and two dynamic cycles for the non-perforated EVG case, versus a single cycle and linear trend observed in the perforated case, thus six instances are plotted for $\phi$ = 0.25.

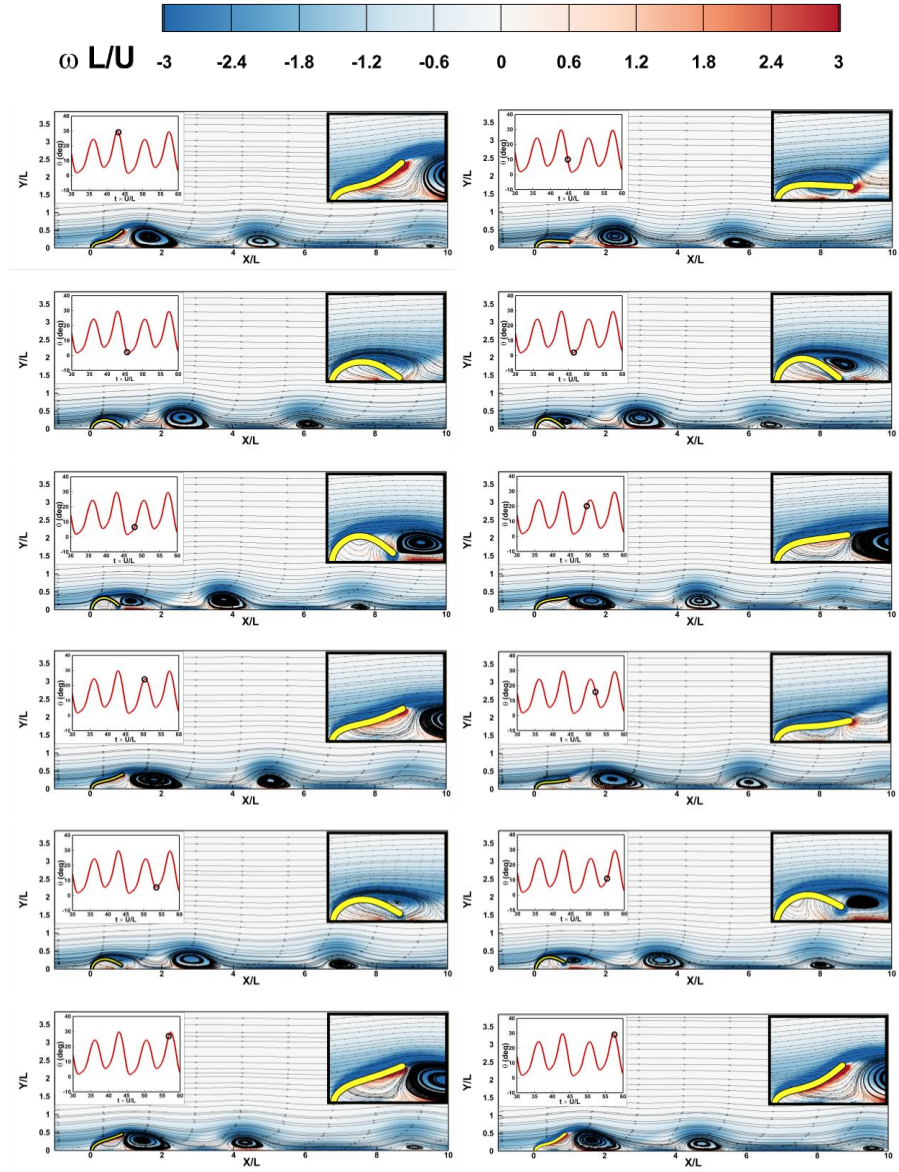

**Figure 20:** Dimensionless vorticity and streamlines plot representing wake structures of a self-excited non-perorated EVG ($\phi = 0$) at $\gamma = 0.004$.

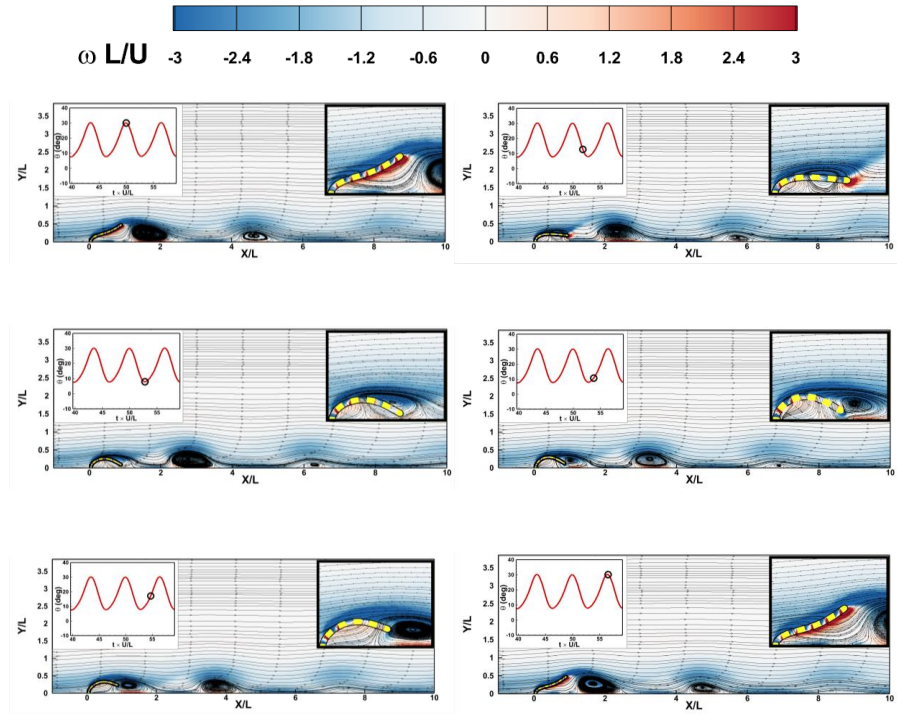

**Figure 21: Dimensionless vorticity and streamlines plot representing wake structures of a self-excited perforated EVG for $\phi = 0.25$ at $\gamma = 0.004$.**

It is evident from Figures 20 and 21 that at the tip of self-excited non-perforated and perforated EVG, the boundary layer becomes unstable and rolls up into vortical structures detaching from the tip region, leading to a periodic shedding of vortices downstream. The non-perforated and perforated EVG responds to the periodic vortex shedding phenomenon, which leads to the oscillation of EVGs. Figures 20 and 21 also illustrate the bleeding effect from the pores in perforated EVGs. As a result of this bleeding flow, the vortical structures in PEVGs are smaller in size and dissipate more quickly compared to those in non-perforated EVGs. It is important to note that the observed bleeding flow and near-wake topology pertain specifically to the pore layout considered in this study. Variations in pore layout, as reported in the literature [100,101], can influence these results, although the effect of pore layout is of secondary importance compared to the impact of the level of perforation [100].

### 6.6. Lock-in analysis

Depending on the ratio of the oscillation frequency induced by vortex shedding to the EVG's natural frequency, the EVG oscillation can result in a lock-in phenomenon, leading to large responses. The lock-in phenomenon occurs not only when the vortex shedding frequency matches the natural frequency of the structure, but it can also occur if the frequency ratio is not exactly one, as demonstrated theoretically using the nonlinear Van der Pol oscillator [102]. The range of frequency ratios that lead to lock-in primarily depends on the magnitude of the fluctuating lift force. As the fluctuating lift force increases, the width of the lock-in region also expands, allowing the system to experience the lock-in phenomenon over a broader range of frequencies.

To check whether EVG experiences lock-in phenomenon, Figure 22 plots the dimensionless frequency of oscillation (vortex shedding frequency) and natural frequency of EVGs. The frequency of oscillation is extracted from the predicted EVG responses. The natural frequency of perforated EVG, clamped at one end and in a dry mode (not immersed in flow), is given in [103] as follows:

$$f_{ni} = \frac{1}{2\pi} * \left( \frac{\frac{(EI)_P}{(\rho_S A)_P} * \left(\frac{k_i}{L}\right)^4}{1 + \frac{(EI)_P}{(GA)_P} * \left(\frac{k_i}{L}\right)^2} \right)^{1/2} \tag{12}$$

Where, $(EI)_P$, $(\rho_S A)_P$, and $(GA)_P$ are defined as bending stiffness, mass per unit length and shear stiffness for perforated filaments. Using the definition of $\gamma$ and $\beta$, and assuming that there is negligible shear stiffness, and the added mass for the perforated EVG is similar to the one for non-perforated EVG reported in [64], the wet natural frequency can be expressed as:

$$f_{ni} = \frac{k_i^2}{2\pi} \sqrt{\frac{\gamma * K}{\beta * X + C_m * \frac{\pi}{4}}} \tag{13}$$

where,

$$X = \frac{[1 - N(B-2)]B}{N+B} \tag{14}$$

Here, $f_{ni}$ is the ith-order dimensionless natural frequency, $K$ is the coefficient described in Equation 11 showing the effect of porosity, and $C_m$ is the added mass coefficient. It should be noted that the added mass is generally a function of perforation. Even though studies in the literature have shown the effect of perforation on the added mass[104-115], these studies have investigated configurations that differ significantly from the conditions used in this study-including different geometry of the structure; movement of the structure such as heave motion; perforation ratio; layout and configuration of the pores, and level of oscillation amplitude of the structure. Thus, in this study, we assume it as constant $C_m = 1$, which is similar to that of the non-perforated case, as reported in studies for non-perforated structures [64,116]. The coefficients of the first and second natural frequencies are $k_1 = 1.875$ and $k_2 = 4.694$ respectively. It is worth noting that both the constants K and X go to one for non-perforated case (N=0), therefore the Equation 13 reduces to the following equation for non-perforated cases which is also reported in [64].

$$f_{ni} = \frac{k_i^2}{2\pi} \sqrt{\frac{\gamma}{\beta + C_m * \frac{\pi}{4}}} \tag{15}$$

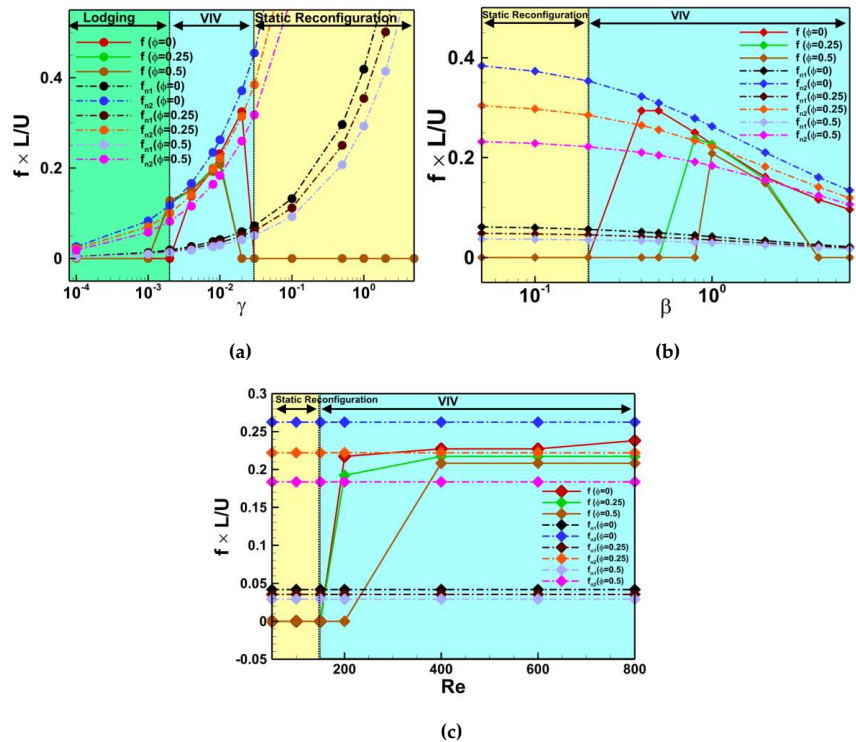

**Figure 22: First and second natural frequency of the self-excited non-perforated and perforated EVG.**

The dimensionless frequency of oscillation (vortex shedding frequency) and natural frequency of EVGs is plotted for the entire range of $\gamma$, $\beta$ and Re in Figures 22a, b and c respectively. Figure 22a shows that, for the cases in VIV mode, the oscillation frequency of the self-excited non-perforated EVG consistently ranges from 0.14 to 0.32. This observed range of dimensionless frequencies in VIV mode agrees well with the Strouhal number for vortex shedding over a bluff body in literature (which is around 0.2). Moreover, it is also observed that the oscillation frequency of the self-excited non-perforated EVG is close to the second natural frequency $f_{n2}$ and far away from the first natural frequency $f_{n1}$ of the self-excited non-perforated EVG, suggesting that EVG oscillation can be locked onto the second natural frequency of the self-excited non-perforated EVG for the considered test conditions. For the cases with porosity (at $\phi$ =0.25 and $\phi$ =0.5), the natural frequency of EVG reduces as shown in Equation 13 and Figure 22a. Therefore, the lock-in occurs at lower frequency, shifting VIV region toward lower $\gamma$ as shown in Figure 11.

Figure 22b indicates that, in VIV mode, the oscillation frequency of the self-excited non-perforated EVG falls between 0.1 and 0.32, close to the range observed in Figure 22a. Again, the oscillation frequency of the self-excited non-perforated EVG aligns closely with its second natural frequency $f_{n2}$. With an increase in porosity (at $\phi$ =0.25 and $\phi$ =0.5), the natural frequency drops, shifting lock-in for perforated EVGs to higher $\beta$ values where frequencies reduce. However, at very high $\beta$ ($\beta > 3$), perforated EVGs experience lodging instead of VIV since there is not enough lift for perforated EVG at such conditions to counteract the large bending moment on EVGs due to its weight, unlike non-perforated cases. This shrinks significantly the range of $\beta$ leading to lock-in and VIV for perforated cases.

According to Figure 22c, for cases in VIV mode, the oscillation frequency of the self-excited non-perforated EVG remains within the range of 0.22 to 0.24, which is again in agreement with the Strouhal number for vortex shedding over a bluff body as documented in the literature. Additionally, the oscillation frequency of the self-excited non-perforated EVG is again near the second natural frequency $f_{n2}$. With an increase in porosity of $\phi$ =0.25, the EVG still shows VIV at similar Re number that the non-perforated EVG does. However, at very high porosity of $\phi$ =0.5, VIV region shifts to higher Re number.

## 7. Conclusion

We conducted high-fidelity 2-way coupled Fluid-Structure Interaction (FSI) simulations focusing on a novel perforated self-excited vortex generator (PEVG). The responses of PEVGs were investigated across a wide range of dimensionless parameters, including dimensionless rigidity, mass ratios, Reynolds numbers, and porosity levels, and the results were compared with those of non-perforated EVGs.

Three different modes—Lodging, VIV (Vortex-Induced Vibration), and static configuration—were observed for PEVGs, similar to those seen in non-perforated EVGs. Despite these similarities, it was found that the amplitude of oscillation for PEVGs is lower compared to non-perforated EVGs, and the range of VIV is shifted in the case of PEVGs. This shift in the VIV region is attributed to the reduced natural frequency of the perforated self-excited EVGs compared to non-perforated self-excited EVGs. The results showed that the differences between the responses of perforated and non-perforated VGs are partly due to changes in the mechanical properties of perforated EVGs, as the moment of inertia varies along the length of the perforated EVGs, and partly due to the effect of pores on fluid loads. The study on loads also revealed that the drag force is often less for perforated cases, making them suitable as vortex generators. They often cause less pressure drop while still exhibiting all the response modes observed in non-perforated EVGs to enhance mixing. Additionally, the investigation of the lock-in phenomenon showed that EVG oscillations are locked onto the second natural frequency for all non-perforated and perforated EVGs experiencing VIV for considered test conditions, resulting in an EVG profile with a full sinusoidal wave along its length. Harmonic analyses revealed the presence of higher harmonics in non-perforated EVGs, but these are not as prominent as the first peak and are significantly reduced in perforated cases.

Local flow studies demonstrates that at the tip of both non-perforated and perforated EVGs, the boundary layer becomes unstable, leading to periodic vortex shedding downstream and causing EVG oscillation. The results also showed that the bleeding effect from the pores in perforated EVG cases leads to smaller vortical structures that dissipate more quickly compared to those in non-perforated EVGs.

### Data Availability Statement

The data that support the findings of this study are available from the corresponding author upon reasonable request.